\documentclass[fleqn,usenatbib]{mnras}

\usepackage{newtxtext,newtxmath}

\usepackage[T1]{fontenc}

\DeclareRobustCommand{\VAN}[3]{#2}
\let\VANthebibliography\thebibliography
\def\thebibliography{\DeclareRobustCommand{\VAN}[3]{##3}\VANthebibliography}

\usepackage{graphicx}	
\usepackage{amsmath}	
\usepackage{bm}

\title[RSGs in the other side of our Galaxy]{New Red Supergiant Stars in the other side of our Galaxy}

\author[Lin Zhang et al.]{
Lin Zhang, $^{1}$
Bingqiu Chen, $^{1}$\thanks{E-mail: bchen@ynu.edu.cn}
Yi Ren,$^{2}$\thanks{E-mail: yiren@qlnu.edu.cn}
Zehao Zhang,$^{3,4}$
Jian Gao$^{3,4}$
and Biwei Jiang$^{3,4}$
\\
$^{1}$South-Western Institute for Astronomy Research,
Yunnan University, Kunming 650500, P.R. China\\
$^{2}$College of Physics and Electronic Engineering,
Qilu Normal University,
Jinan 250200, P.R. China\\
$^{3}$Institute for Frontiers in Astronomy and Astrophysics,
 Beijing Normal University,
Beijing 102206, P.R. China\\
$^{4}$Department of Astronomy,
Beijing Normal University,
Beijing 100875, P.R. China
}

\date{Accepted XXX. Received YYY; in original form ZZZ}

\pubyear{\the\year{}}

\begin{document}
\label{firstpage}
\pagerange{\pageref{firstpage}--\pageref{lastpage}}
\maketitle

\begin{abstract}
Red supergiant stars (RSGs) are massive stars in a late stage of evolution, crucial for understanding stellar life cycles and Galactic structure. However, RSGs on the far side of our Galaxy have been underexplored due to observational challenges. In this study, we introduce a novel method and present a new catalogue comprising 474 RSGs situated on the far side of the Milky Way, sourced from the OGLE-III catalogue of Variable Stars (OIII-CVS). The identification of these RSGs was made possible by analyzing the granulation parameters extracted from OGLE $I$-band time-series data and the stellar parameters from Gaia DR3. Additionally, we estimate the distances to these RSGs using an empirical relation between their characteristic amplitude, absolute magnitude, and intrinsic color, achieving a distance uncertainty of 13\%. These newly identified RSGs are distributed at Galactocentric distances between 0 and 30\,kpc, and reside roughly 1 to 4\,kpc above and below the Galactic plane. This distribution provides new insights into the structure of the Milky Way, particularly at its outer boundaries. Our results reveal that the vertical distribution of these RSGs is consistent with the flare structure of the Galactic disk, confirming that the far side of the Milky Way exhibits a similar flaring pattern to the near side. This catalogue offers a valuable resource for future detailed studies of RSGs and contributes to a more comprehensive understanding of the Galactic structure and stellar evolution. 
\end{abstract}



\section{Introduction}

Red supergiant stars (RSGs) represent a critical phase in the life cycle of massive stars, which ultimately end in Type II supernovae \citep{Massey2008, Massey2016, VanDyk2017, Levesque2017}. These stars provide essential insights into the mass range and evolutionary pathways of massive stars, as well as the role of metallicity in stellar evolution \citep{Massey2002, Massey2013}. RSGs are also valuable for calibrating period-luminosity (PL) relations, which can be used as standard candles to estimate extragalactic distances \citep{Feast1980, Jurcevic2000, Kiss2006, Yang2011, Yang2012, Soraisam2018, Ren2019}. Additionally, RSGs have a significant impact on both stellar evolution and the interstellar medium. Their substantial mass loss influences the final stages of stellar evolution \citep{Beasor2018} and contributes to the enrichment of the interstellar environment \citep{Wen2024}. Therefore, constructing a comprehensive and uncontaminated sample of RSGs is crucial for understanding their properties and broader effects. 

Traditionally, RSGs are identified by their association with OB clusters or associations at optical wavelengths \citep{Humphreys1978, Garmany1992, Stencel2009}. While catalogues often focus on bright, late-type stars in the vicinity of clusters, it has been observed that only about 2\% of RSGs are cluster members \citep{Messineo2017}. Another method for identifying RSGs involves their placement on the Hertzsprung-Russell (color-magnitude) diagram. In extragalactic systems, this approach is often hindered by contamination from foreground stars. However, Gaia astrometric data can help remove foreground dwarfs \citep{Yang2019,Ren2021b}, and near-infrared color-color diagrams ($J-H/H-K_{\rm S}$) can distinguish between foreground dwarfs and (super)giant members, utilizing the $H$-band's sensitivity to surface gravity log\,$g$ \citep{Ren2021a}. 

In our own Galaxy, identifying RSGs is particularly challenging due to our location within the Galactic disk. The high and uneven levels of interstellar extinction complicate the measurement of stellar intrinsic luminosity and color, making it difficult to build a complete catalogue of Galactic RSGs. \citet{Gehrz1989} estimated that there could be around 5200 RSGs in the Milky Way. However, fewer than 1000 have been identified to date \citep{Messineo2019}, and many of these detections are contaminated by other evolved stars, such as asymptotic giant branch stars (AGBs) and luminous red giant branch stars (RGBs), due to the difficulty in distinguishing them from RSGs on color-magnitude diagrams (CMDs).

In this study, we aim to distinguish AGBs from RSGs by analyzing the variability characteristics from the Optical Gravitational Lensing Experiment (OGLE; \citealt{Udalski1997, Udalski2003, Soszy2013, Udalski2015}) light curves, and exclude RGBs using stellar parameters from Gaia DR3 \citep{Gaia2016,Gaia2023}. The use of light variation in distinguishing AGBs and RSGs is rooted in the relations between variability parameters and stellar parameters, which helps determine the stellar intrinsic luminosity and color index, thereby identifying high-luminosity, low-effective-temperature RSG candidates. For instance, the PL relation in RSG pulsations can be utilized to identify candidates that meet the luminosity criteria for RSGs. However, light variation of RSGs is attributed to both pulsation and surface granulation evolution, with granulation signals introducing irregularities in the variability \citep{Kiss2006}. Typically, only those RSGs with relatively regular variability can have their periods accurately determined, making it difficult to obtain a complete sample. Conversely, the determination of granulation characteristic parameters (characteristic timescale, $\tau_{\rm{gran}}$ and characteristic amplitude, $\sigma_{\rm{gran}}$) does not require classifying light curves. Since there are scaling relations between granulation parameters and stellar parameters \citep{Ren2020,Zhang2024}, once the granulation parameters are determined, they can be used to exclude the contamination of those AGB stars.

RSGs are excellent tracers of the Galactic structure. However, our understanding of their spatial distribution is limited by sample incompleteness and uncertain distances. \citet{Hey2023} used kernel density estimation (KDE) to model the period-amplitude-luminosity space and measured distances to long-period variables (LPVs) in the Galactic bulge using OGLE data. However, the approach only determines the distances to stars that are located in regions with high stellar density. For RSGs with small sample sizes, the distances are inaccurate using the method of \citet{Hey2023}. In this work, we determine the distances of RSGs using the scaling relation.

\section{Data}
Our work is based on the time-series data from the OGLE and near-infrared photometric data from the Two Micron All Sky Survey (2MASS, \citealt{Skrutskie2006}). 

The OGLE survey has been regularly observing the Galactic Bulge, Galactic disk and Magellanic Clouds (MCs) since 1997. The OGLE observations are conducted using the 1.3-meter telescope located at Las Campanas Observatory in Chile. The survey data consist of Johnson $V$-band and Cousins $I$-band photometry. In this work, we focus on time-series data from the third phase of the OGLE observation (OGLE-III) $I$-band, which spans nearly a decade, from 2001 to 2010, covering the Galactic bulge. These observations have a cadence from 3 to 8 days, with most sources observed 600 to 800 times. In addition, the typical photometric uncertainty is about 0.005\,mag, enabling the detection of detailed stellar variability features. We search for RSGs from the fifteenth part of OGLE-III catalogue of Variable Stars (OIII-CVS), which provides 232,406 LPVs located in the Galactic bulge based on the second (OGLE-II; \citealt{Udalski1997}) and the third (OGLE-III; \citealt{Udalski2003}, \citealt{Soszy2013}) phases of the OGLE survey.

The near-infrared photometric data are obtained from the 2MASS survey, conducted in the $J$, $H$, and $K_{\rm S}$ bands using two 1.3-meter telescopes, respectively, located at Mt. Hopkins and CTIO. The 2MASS photometric precision is less than 0.03\,mag for bright sources.

To provide a criterion for identifying RSG candidates and to later roughly assess the completeness and purity of the RSGs identification, we selected known RSG samples and non-RSG samples as the so-called reference sample. The reference sample are obtained from Table 2 of \citet{Zhang2024}, who further refined the reliable RSGs based on the RSG sample from \cite{Ren2021b} by using radial velocity from Gaia DR3 \citep{Katz2023} and the Apache Point Observatory Galactic Evolution Experiment (APOGEE, \citealt{Abdurro2022}) and the Gaia CMD. The non-RSGs are taken from the CMD of \cite{Ren2021b}, which consist mainly of AGBs and RGBs. Finally, the reference sample stars contain 2309 RSGs and 11,352 non-RSGs in the Large Magellanic Cloud (LMC), and 1210 RSGs and 3476 non-RSGs in the Small Magellanic Cloud (SMC). It's worth noting that the non-RSGs with time-series data are mainly the AGBs. Of the 14,828 non-RSGs, we identify 12,702 of them as AGBs based on the $(J-K_{\rm S})_{0}/K_{\rm S 0}$ diagram \citep{Ren2021b}. We collect the $I$-band light curves from the OGLE-III for the reference sample stars, observed between 2001 and 2010.

\section{Method}
\subsection{Determination of granulation parameters} 

\begin{figure}
\includegraphics[width=0.42\textwidth, height=6.3cm]{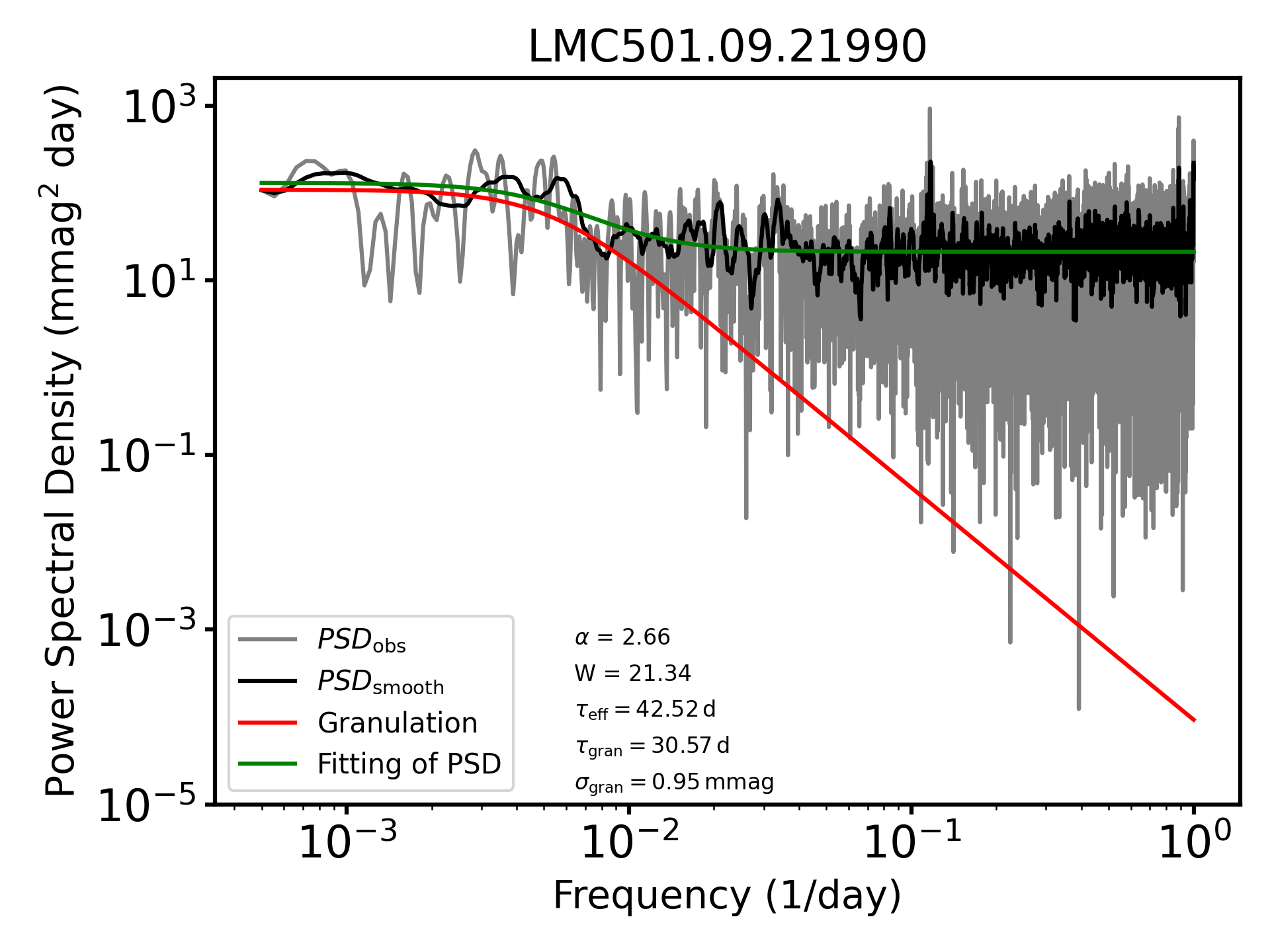}
\caption{Example of PSD fitting for selected RSG LMC501.09.21990. The gray line represents the derived PSD using the Lomb-Scargle \citep{Lomb1976, Scargle1982} method, while the black line is the smoothed PSD. The green line shows the fitting of smoothed PSD using Equation~(\ref{equ:OCT}). The red line represents the fitting of Equation~(\ref{equ:OCT}) without white noise. The fitting parameters are listed in the figure.}
\label{fig:LMC501.09.21990}
\end{figure}

The granulation parameters, which are mainly characterized by two parameters, the characteristic timescale and the characteristic amplitude, can be used to effectively eliminate the AGBs contamination. The granulation parameters can be obtained by fitting the background of the power spectrum. The model for granulation parameters was first proposed by \cite{Harvey1985}. Subsequently, \cite{Mathur2011} and \cite{Ren2020} fitted the background of the power spectral density (PSD) of RGBs and RSGs using the modified Harvey functions, the COR function \citep{Mosser2009, Mosser2011}, to extract the granulation parameters. Similarly, we use the Harvey-like function, specifically the Octave (OCT) function \citep{Hekker2010}, to fit the background of the PSD of the reference sample stars and capture their granulation parameters. The OCT function has the following form:
\begin{equation}
P(\nu) = W + \frac{4 \sigma_{\rm gran}^2 \tau_{\rm gran}}{1 + (2 \pi \nu \tau_{\rm gran}) ^ \alpha}, 
\label{equ:OCT}
\end{equation}
where $W$ represents the white noise and $\alpha$ is a positive parameter that represents the slope of the decay. Fig.~\ref{fig:LMC501.09.21990} shows an example of fitting the PSD background to obtain the granulation parameters.

To compare characteristic timescales at different values of $\alpha$, \cite{Mathur2011} defined an effective timescale $\tau_{\rm{eff}}$, as the e-folding time of the autocorrelation function (ACF). The ACF is calculated from the PSD numerically. Hence, we transform $\tau_{\rm{gran}}$ to $\tau_{\rm{eff}}$ based on the ACF. The granulation parameters as well as their uncertainties are calculated using the Markov Chain Monte Carlo (MCMC) method.

\subsection{Identification of RSG candidates} 

Due to the inadequate sampling of the light curve, which may cause a great uncertainty on the granulation timescales, we discard the light curves of reference sample stars with less than 50 detections. For the remaining of light curves, we calculate the granulation parameters and their uncertainties. Only sources with error smaller than 50 days in $\tau_{\rm{eff}}$ and 50\,mmag in $\sigma_{\rm{gran}}$ are remained, including 1146 RSGs and 8920 non-RSGs. There are about 7889 AGBs in the non-RSGs.

\begin{figure}
\centering
\hspace{-10mm}
\includegraphics[width=0.41\textwidth, height=6.3cm]{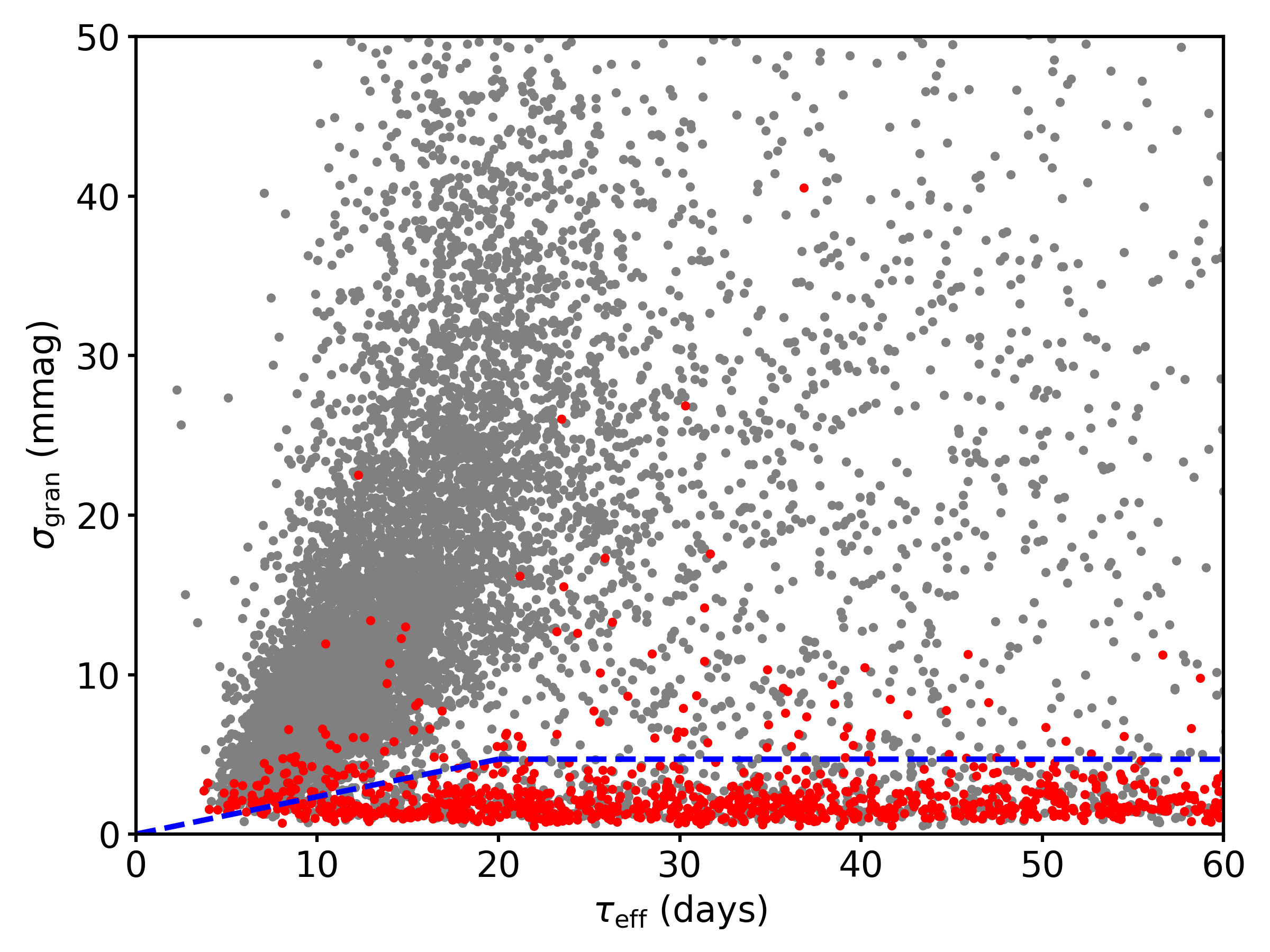}
\includegraphics[width=0.5\textwidth, height=6.3cm]{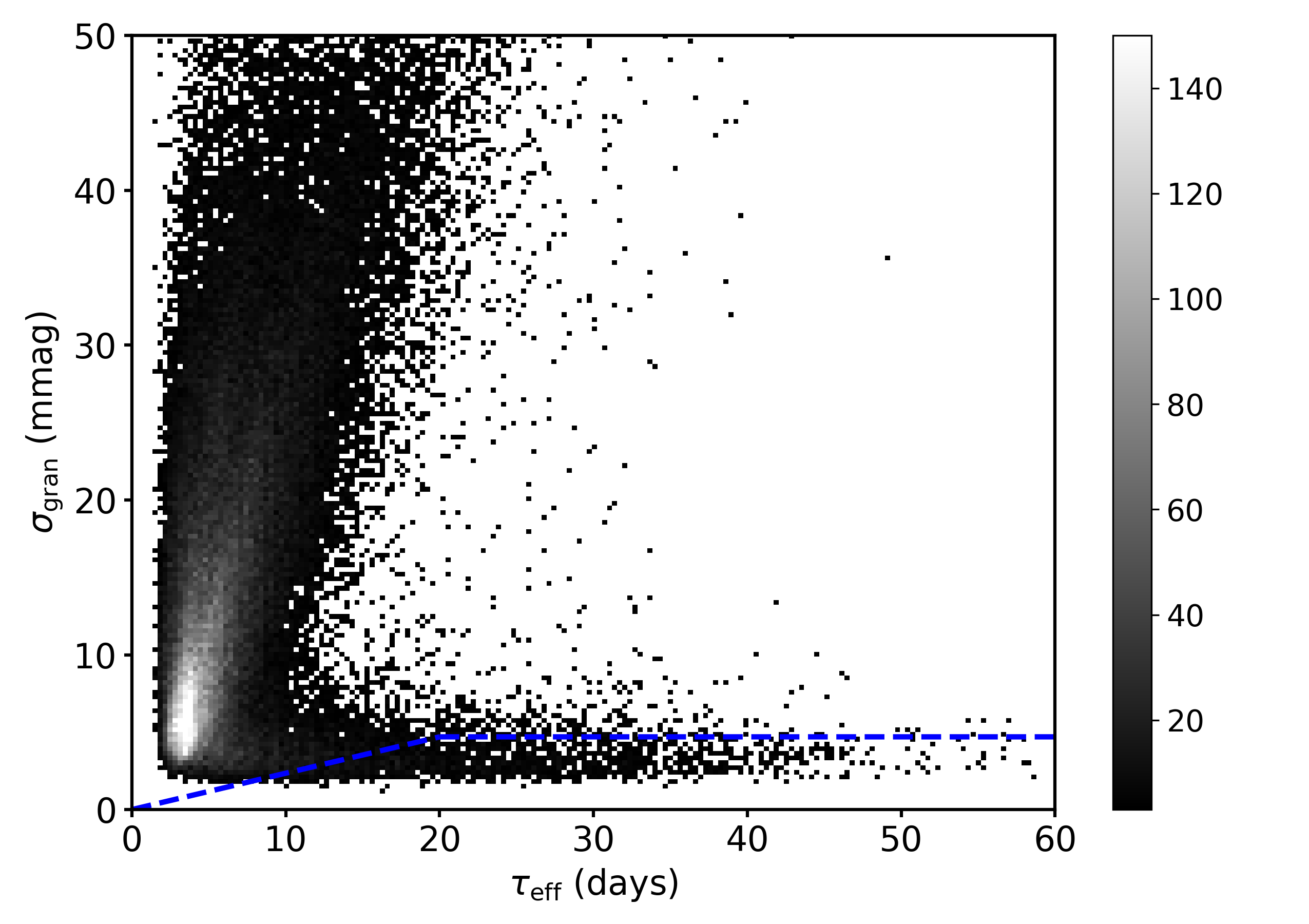}
\caption{Distribution of the granulation parameters for RSGs and non-RSGs in the Magellanic Clouds (upper panel) and that for OIII-CVS candidate stars (lower panel). In the upper panel, the grey dots represent the non-RSGs, while the red dots represent the RSGs. The visual division between RSGs and non-RSGs are indicated by the blue lines.}
\label{fig:training_sample}
\end{figure}

We present the distribution of granulation parameters for the reference sample stars in the upper panel of Fig.~\ref{fig:training_sample}. It appears that distinguishing non-RSGs in the $\tau_{\rm{eff}}-\sigma_{\rm{gran}}$ diagram is relatively straightforward. For RSGs, the $\tau_{\rm{eff}}$, have a wide range, from several to a few dozen days, and $\sigma_{\rm{gran}}$, are primarily distributed around several mmag. To balance the precision and recall, we manually inspected the division of RSG candidates and non-RSGs. The boundaries were determined as follows:

\begin{equation}
\begin{aligned}
    \sigma_{\rm{gran}} &= 4.7 \quad  ( 0 \leq \tau_{\rm{eff}} \textless 20), \\
    \sigma_{\rm{gran}} &= \frac{47}{200}\tau_{\rm{eff}} \quad (\tau_{\rm{eff}} \geq 20).
\label{equ:boundary}
\end{aligned}
\end{equation}

Before applying the boundary lines to the LPVs from OIII-CVS for selecting RSG candidates, we preprocess these LPVs in the same way as the reference sample stars, specifically, constraining the detections and the uncertainties of the granulation parameters. These cuts leave approximately 13,600 LPVs as the candidate sample stars for selecting RSG candidates. Then, we apply the dividing lines, as shown in Equation~(\ref{equ:boundary}), to the candidate sample stars. The lower panel of Fig.~\ref{fig:training_sample} shows the density distribution of the candidate sample stars and the dividing lines in the $\sigma_{\rm{gran}}$ and $\tau_{\rm{eff}}$ space. After eliminating the main source of contamination, AGBs, a total of 5403 RSG candidates are identified. 

\begin{figure}
\centering
\includegraphics[width=0.4\textwidth]{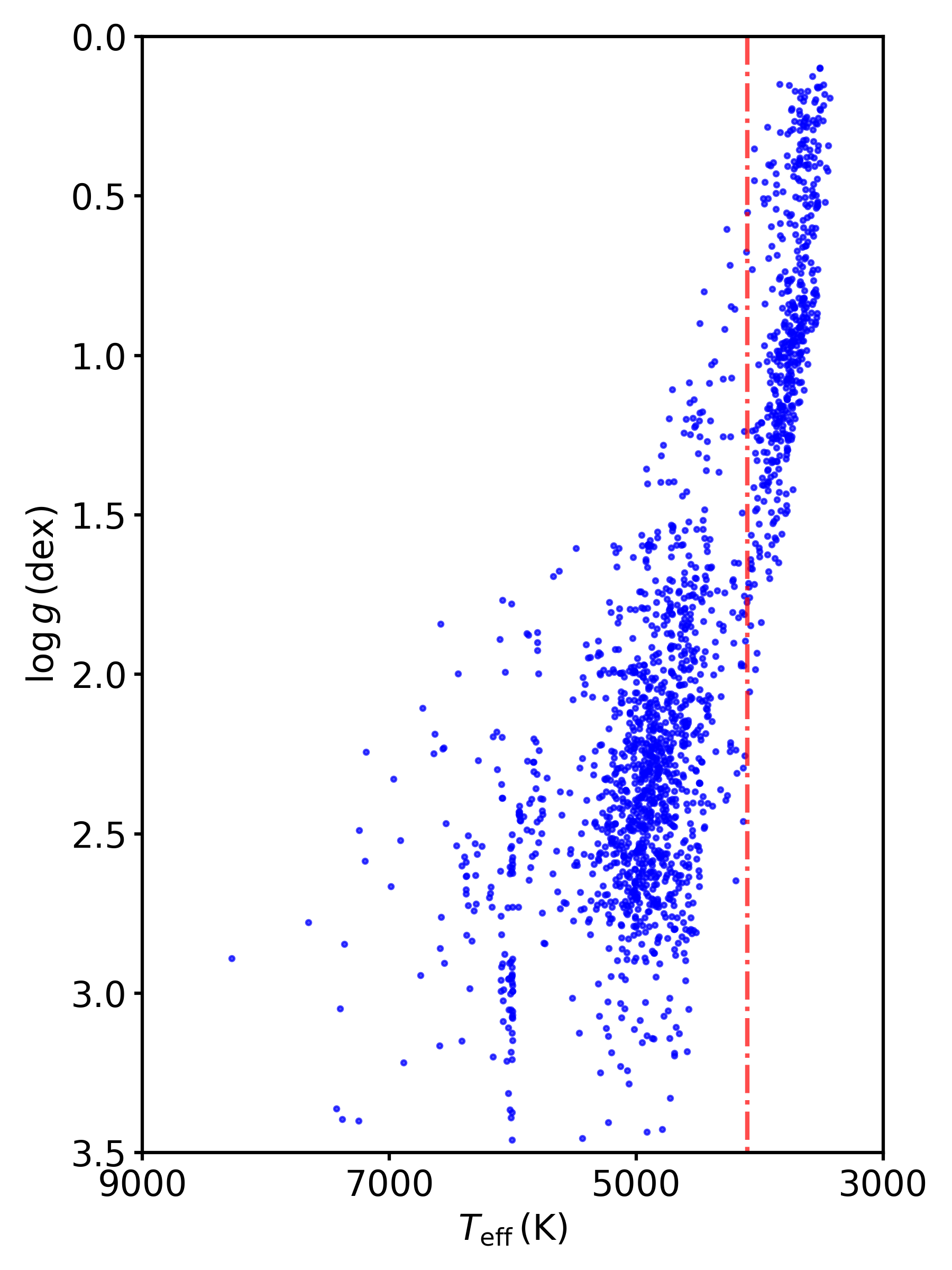}
\caption{Distribution of the Gaia DR3 $T_{\rm eff}$ and log\,$g$ for 1967 RSG candidates. The visual division between RSGs and RGBs is indicated by the red dashed-dotted line.}
\label{fig:exculding_rgb}
\end{figure}

In this candidate sample, another source of contamination comes from luminous RGB stars. The tip of the RGBs overlaps significantly in luminosity with the faint RSGs examined in this study. Due to their similar luminosities and effective temperatures ($T_{\rm eff}$), it is challenging to distinguish RSGs from RGBs based solely on granulation parameters. To address this issue, we apply a method similar to that used by \citet{Messineo2023}, utilizing Gaia stellar parameters to eliminate RGBs. 

We cross-matched the 5403 RSG candidates identified in the previous step with Gaia DR3, resulting in 5013 common sources. Of these, 1967 stars have stellar parameters derived from Gaia's photometric and BP/RP XP spectral data \citep{Fouesneau2023}. Fig.~\ref{fig:exculding_rgb} shows the distribution of these stars in terms of their $T_{\rm eff}$ and surface gravity (log\,$g$). The Kiel diagram (log\,$g$ versus $T_{\rm eff}$) reveals no visible contamination from AGB stars, indicating their successful removal using granulation parameters. The candidates split into two distinct groups: one in the lower left, corresponding to RGBs, with $T_{\rm eff}$ ranging from 4100 to 5500\,K and log\,$g$ between 1.0 and 3.0\,dex; and another in the upper right, representing the RSGs we are targeting. It should be noted that the temperature estimates provided by \citet{Fouesneau2023} are somewhat lower than the actual values for RSGs. This discrepancy arises because their study lacks sufficient RSG templates, and RSGs often have circumstellar dust \citep{Wen2024}, which affects temperature measurements. As a result, RSGs appear slightly offset from RGBs in Fig.~\ref{fig:exculding_rgb}. Thanks to this, our approach allows for the identification of accurate and reliable RSG candidates. From this process, we select 474 sources with $T_{\rm eff}$ below 4100\,K as our final RSG sample. To check the robustness of our selection, we also used the stellar parameters from \citet{Zhang2023} and \citet{Andrae2023} to plot the identified RSGs on the Kiel diagram. It is found that the identified RSGs are located in the low-temperature, low-surface-gravity region of the Kiel diagram, which supports the reliability of our RSG sample selection.

\section{Results} \label{sec:results and discussion}

In this study, we have identified 474 new RSG candidates from the OIII-CVS and have estimated their distances. The determination of the distance and the completeness, purity, and limitations of our catalogue are discussed in detail in the appendix. Our methodology is effective primarily for identifying faint RSGs, whose characteristic amplitudes are significantly smaller than those of AGB stars. The precision and recall of our sample are above 62\% and 84\%, respectively. Additionally, the distance precision of the derived values is addressed in the appendix, where we report a typical relative distance uncertainty of approximately 13\%.

\begin{figure}
\includegraphics[width=\columnwidth]{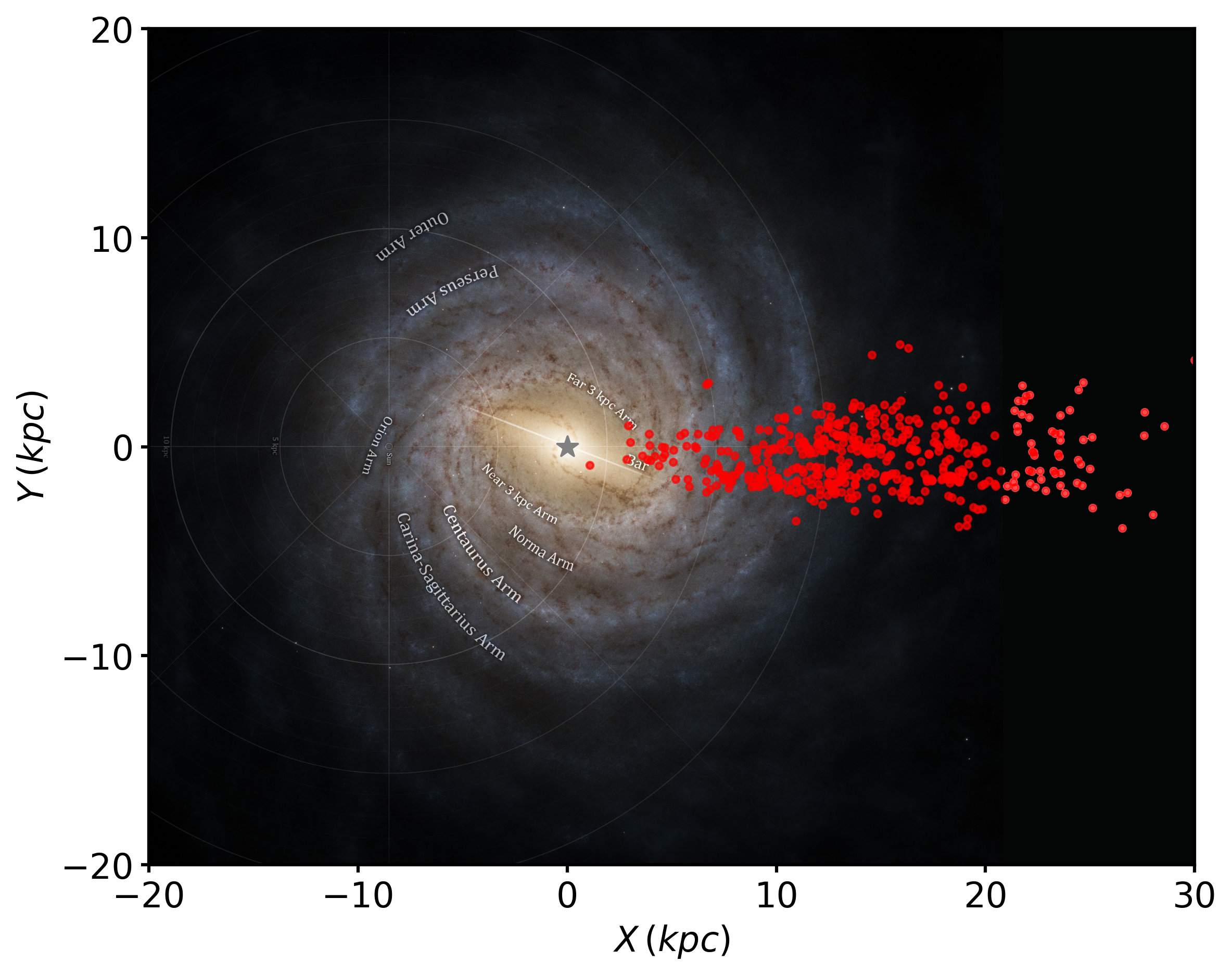}
\caption{The distribution of the new RSG sample in the $X-Y$ plane. The position of the Galactic centre is marked by the grey star symbol. The background image shows the face-on view of the Milky Way's structure (based on Gaia DR3; Credits: ESA/Gaia/DPAC, Stefan Payne–Wardenaar, CC BY-SA 4.0 IGO).}
\label{fig:kde_galaxy}
\end{figure}

We have also examined the three-dimensional spatial distribution of the newly identified RSGs within the Milky Way. Fig.~\ref{fig:kde_galaxy} illustrates the distribution of these RSG candidates in the Galactic plane, overlaid on a map of the face-on view of the Galaxy. The distribution in the $X-Y$ plane reveals that the majority of the RSGs are located at a Galactocentric distance of around 10\,kpc. Some of these newly identified RSGs may align with the Scutum-Centaurus and Sagittarius arms of the Galaxy, suggesting that these stars may trace the structure of the spiral arms. Nevertheless, further precise distance measurements are necessary to substantiate this hypothesis.

\begin{figure}
\includegraphics[width=\columnwidth]{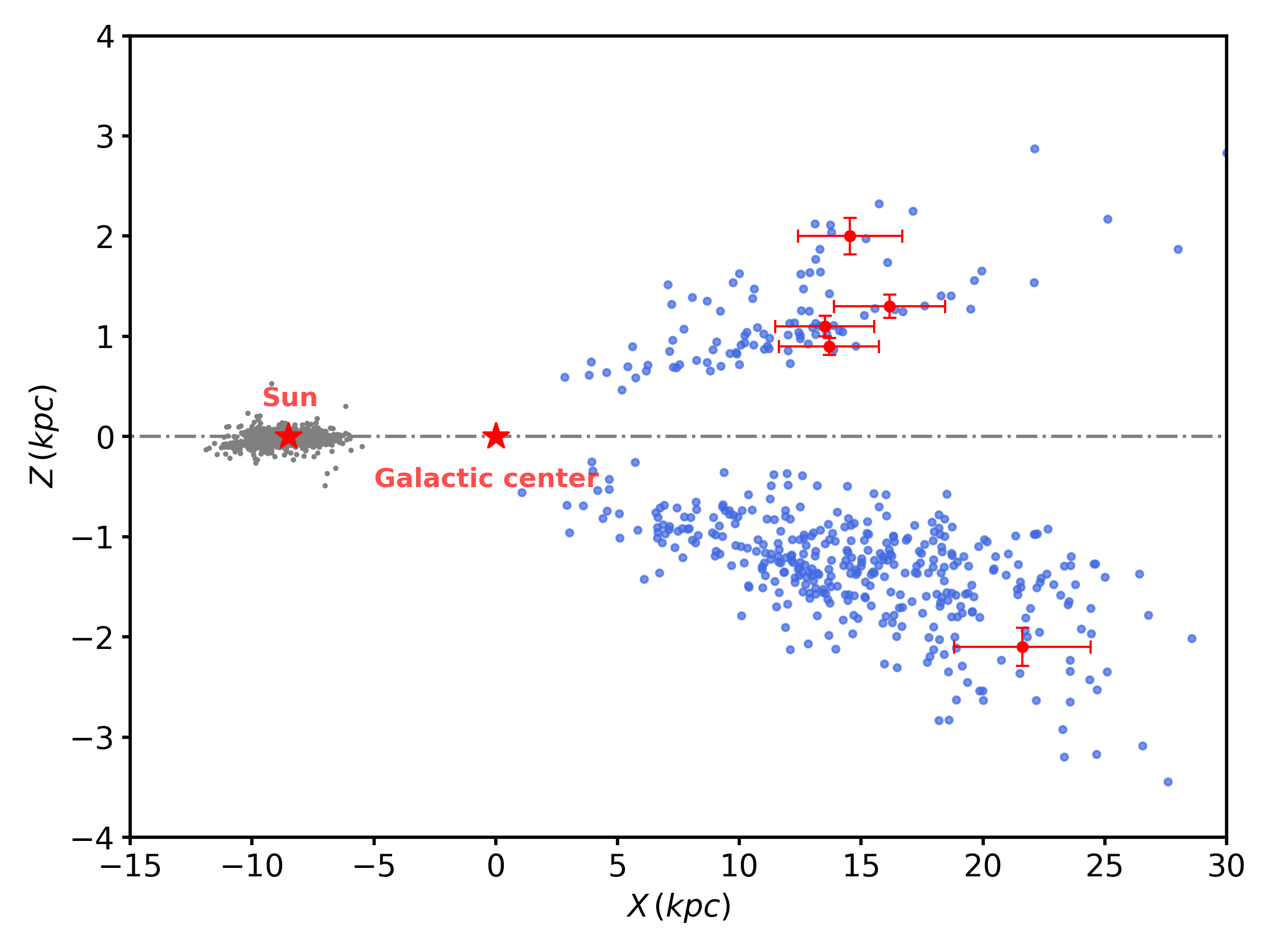}
\caption{The distribution of the RSGs within 30\,kpc is compared to the positions of the Cepheids (red dots) in the $X-Z$ plane. The Galactic centre is located at ($X = 0$\,kpc and $Z = 0$\,kpc) and marked by red stars, while the Sun is also presented by red stars at ($X = - 8.5$\,kpc and $Z = 0$\,kpc). The new RSG candidates are shown in blue dots. The red dots with error bars are Cepheids from \citet{Feast2014} and the grey dots are the RSG candidates from \citet{Messineo2019}. Both datasets are used for comparison.}
\label{fig:flare}
\end{figure}

In terms of their radial distribution, these RSGs are spread across the far side of the Galaxy, with Galactocentric distances ranging from $\sim$ 0 to $\sim$ 30\,kpc. Although we have detected a few RSGs beyond 30\,kpc, we do not include them in the plot due to their sparse numbers and potential contamination by non-RSG sources. The Galactic disk exhibits a cutoff at a Galactocentric distance of approximately 30\,kpc \citep{Sesar2010, Mosenkov2021}. Our results follow this perspective, suggesting that the disk extends to at least 30\,kpc. However, because we have not assessed the completeness of distant RSGs in this work, we cannot definitively establish a sharp cutoff at 30\,kpc. Nevertheless, the discovery of these distant RSGs provides valuable insights into the outer boundaries of the Milky Way.

We also present the distribution of the new RSG candidates in the $X-Z$ plane, as shown in Fig.~\ref{fig:flare}. The majority of these RSGs are located between 1 and 4\,kpc above and below the Galactic plane. The absence of RSGs along the plane itself ($-1^{\circ} \leq b \leq 1^{\circ}$) is due to the heavy dust obscuration in this region, which makes it nearly impossible to detect RSGs in the OGLE $I$ band. Furthermore, given that RSGs are relatively young stars, their lower numbers closer to the Galactic centre are expected. Previous studies typically identified RSGs in the solar neighborhood. The grey dots in Fig.~\ref{fig:flare} represent RSG candidates identified by \citet{Messineo2019}, most of which are located near the Sun. In contrast, the RSG candidates identified in this study are situated on the far side of the Galaxy and are more widely dispersed.

When examining the distribution of new RSGs along the $Z$ direction, we observe a strong flaring pattern. \citet{Feast2014} reported that Cepheid variables are in the outer regions of a flared disk. Thus, we compare the vertical distribution of our sample stars with their five Cepheids, as shown in Fig.~\ref{fig:flare}. The distribution of our sample stars is consistent with that of Cepheids, further illustrating that the RSGs are associated with the flare.

\section{Conclusion} \label{sec:conclusion}
According to granulation parameters extracted from the OGLE $I$ band time-series data and the stellar parameters from Gaia DR3, we present a total of 474 new RSG candidates.  We believe this catalogue is reliable, as the selection method is not affected by extinction or color. However, the method is limited to select the faint RSGs.

Additionally, we derive the distances to the new RSGs using the scaling relation based on absolute magnitude in $K_{\rm S}$-band, intrinsic color and $\sigma_{\rm gran}$, with a dispersion of 13\%. We present the distribution of RSG candidates in the $X-Y$ plane, suggesting a possible association between the new RSGs and the Scutum Centaurus and Sagittarius arms. Moreover, the positions of the newly discovered RSGs expand our understanding of the distribution of RSGs. The distances of RSG candidates are much farther than those previously identified. We also show the distribution of the new RSG sample in the $X-Z$ plane, revealing a trend consistent with a flared structure.

This catalogue is likely to serve as a foundation for future studies on the evolution of massive stars or the structure of the Galaxy. In future work, we may explore the potential of using the method to search for RSGs in extragalactic systems and confirm extragalactic distances.

\section*{Acknowledgements}

This work is supported by the National Natural Science Foundation of China 12173034, 12322304 and 12203025, National Natural Science Foundation of Yunnan Province 202301AV070002, the Xingdian talent support program of Yunnan Province, Shandong Provincial Natural Science Foundation through project ZR2022QA064 and Shandong Provincial University Youth Innovation and Technology Support Program through grant No. 2022KJ138. We acknowledge the science research grants from the China Manned Space Project with NO. CMS-CSST-2021-A09, CMS-CSST -2021-A08, and CMS-CSST -2021-B03. 

This work has made use of data from the European Space Agency (ESA) mission {\it Gaia} (\url{https://www.cosmos.esa.int/gaia}), processed by the {\it Gaia}
Data Processing and Analysis Consortium (DPAC,
\url{https://www.cosmos.esa.int/web/gaia/dpac/consortium}). Funding for the DPAC has been provided by national institutions, in particular the institutions participating in the {\it Gaia} Multilateral Agreement.
This work makes use of data products from the Two Micron All Sky Survey, which is a joint project of the University of Massachusetts and the Infrared Processing and Analysis Center/California Institute of Technology, funded by the National Aeronautics and Space Administration and the National Science Foundation. This work also has made use of data from the OGLE.

\section*{Data Availability}
The catalogue of our newly identified RSGs are available in the online supplementary material. We list the OGLE-ID, J2000 equatorial coordinates, 2MASS $JHK_{S}$ magnitudes, distances, distance uncertainties, as well as the granulation parameters in each line of the catalogue.



\bibliographystyle{mnras}
\bibliography{example} 

\begin{thebibliography}{}
\makeatletter
\relax
\def\mn@urlcharsother{\let\do\@makeother \do\$\do\&\do\#\do\^\do\_\do\%\do\~}
\def\mn@doi{\begingroup\mn@urlcharsother \@ifnextchar [ {\mn@doi@} {\mn@doi@[]}}
\def\mn@doi@[#1]#2{\def\@tempa{#1}\ifx\@tempa\@empty \href {http://dx.doi.org/#2} {doi:#2}\else \href {http://dx.doi.org/#2} {#1}\fi \endgroup}
\def\mn@eprint#1#2{\mn@eprint@#1:#2::\@nil}
\def\mn@eprint@arXiv#1{\href {http://arxiv.org/abs/#1} {{\tt arXiv:#1}}}
\def\mn@eprint@dblp#1{\href {http://dblp.uni-trier.de/rec/bibtex/#1.xml} {dblp:#1}}
\def\mn@eprint@#1:#2:#3:#4\@nil{\def\@tempa {#1}\def\@tempb {#2}\def\@tempc {#3}\ifx \@tempc \@empty \let \@tempc \@tempb \let \@tempb \@tempa \fi \ifx \@tempb \@empty \def\@tempb {arXiv}\fi \@ifundefined {mn@eprint@\@tempb}{\@tempb:\@tempc}{\expandafter \expandafter \csname mn@eprint@\@tempb\endcsname \expandafter{\@tempc}}}

\bibitem[\protect\citeauthoryear{{Abdurro'uf} et~al.,}{{Abdurro'uf} et~al.}{2022}]{Abdurro2022}
{Abdurro'uf} et~al., 2022, \mn@doi [\apjs] {10.3847/1538-4365/ac4414}, \href {https://ui.adsabs.harvard.edu/abs/2022ApJS..259...35A} {259, 35}

\bibitem[\protect\citeauthoryear{{Andrae}, {Rix}  \& {Chandra}}{{Andrae} et~al.}{2023}]{Andrae2023}
{Andrae} R.,  {Rix} H.-W.,   {Chandra} V.,  2023, \mn@doi [\apjs] {10.3847/1538-4365/acd53e}, \href {https://ui.adsabs.harvard.edu/abs/2023ApJS..267....8A} {267, 8}

\bibitem[\protect\citeauthoryear{{Beasor} \& {Davies}}{{Beasor} \& {Davies}}{2018}]{Beasor2018}
{Beasor} E.~R.,  {Davies} B.,  2018, \mn@doi [\mnras] {10.1093/mnras/stx3174}, \href {https://ui.adsabs.harvard.edu/abs/2018MNRAS.475...55B} {475, 55}

\bibitem[\protect\citeauthoryear{{Chen}, {Guo}, {Gao}, {Yang}, {Liu}  \& {Jiang}}{{Chen} et~al.}{2022}]{Chen2022}
{Chen} B.~Q.,  {Guo} H.~L.,  {Gao} J.,  {Yang} M.,  {Liu} Y.~L.,   {Jiang} B.~W.,  2022, \mn@doi [\mnras] {10.1093/mnras/stac072}, \href {https://ui.adsabs.harvard.edu/abs/2022MNRAS.511.1317C} {511, 1317}

\bibitem[\protect\citeauthoryear{{Feast}, {Catchpole}, {Carter}  \& {Roberts}}{{Feast} et~al.}{1980}]{Feast1980}
{Feast} M.~W.,  {Catchpole} R.~M.,  {Carter} B.~S.,   {Roberts} G.,  1980, \mn@doi [\mnras] {10.1093/mnras/193.2.377}, \href {https://ui.adsabs.harvard.edu/abs/1980MNRAS.193..377F} {193, 377}

\bibitem[\protect\citeauthoryear{{Feast}, {Menzies}, {Matsunaga}  \& {Whitelock}}{{Feast} et~al.}{2014}]{Feast2014}
{Feast} M.~W.,  {Menzies} J.~W.,  {Matsunaga} N.,   {Whitelock} P.~A.,  2014, \mn@doi [\nat] {10.1038/nature13246}, \href {https://ui.adsabs.harvard.edu/abs/2014Natur.509..342F} {509, 342}

\bibitem[\protect\citeauthoryear{{Fouesneau} et~al.,}{{Fouesneau} et~al.}{2023}]{Fouesneau2023}
{Fouesneau} M.,  et~al., 2023, \mn@doi [\aap] {10.1051/0004-6361/202243919}, \href {https://ui.adsabs.harvard.edu/abs/2023A&A...674A..28F} {674, A28}

\bibitem[\protect\citeauthoryear{{Gaia Collaboration} et~al.,}{{Gaia Collaboration} et~al.}{2016}]{Gaia2016}
{Gaia Collaboration} et~al., 2016, \mn@doi [\aap] {10.1051/0004-6361/201629272}, \href {https://ui.adsabs.harvard.edu/abs/2016A&A...595A...1G} {595, A1}

\bibitem[\protect\citeauthoryear{{Gaia Collaboration} et~al.,}{{Gaia Collaboration} et~al.}{2023}]{Gaia2023}
{Gaia Collaboration} et~al., 2023, \mn@doi [\aap] {10.1051/0004-6361/202243940}, \href {https://ui.adsabs.harvard.edu/abs/2023A&A...674A...1G} {674, A1}

\bibitem[\protect\citeauthoryear{{Garmany} \& {Stencel}}{{Garmany} \& {Stencel}}{1992}]{Garmany1992}
{Garmany} C.~D.,  {Stencel} R.~E.,  1992, \aaps, \href {https://ui.adsabs.harvard.edu/abs/1992A&AS...94..211G} {94, 211}

\bibitem[\protect\citeauthoryear{{Gehrz}}{{Gehrz}}{1989}]{Gehrz1989}
{Gehrz} R.,  1989, in {Allamandola} L.~J.,  {Tielens} A.~G.~G.~M.,  eds,  IAU Symposium Vol. 135, Interstellar Dust. p.~445

\bibitem[\protect\citeauthoryear{{Harvey}}{{Harvey}}{1985}]{Harvey1985}
{Harvey} J.,  1985, in {Rolfe} E.,  {Battrick} B.,  eds,  ESA Special Publication Vol. 235, Future Missions in Solar, Heliospheric \& Space Plasma Physics. p.~199

\bibitem[\protect\citeauthoryear{{Hekker} et~al.,}{{Hekker} et~al.}{2010}]{Hekker2010}
{Hekker} S.,  et~al., 2010, \mn@doi [\mnras] {10.1111/j.1365-2966.2009.16030.x}, \href {https://ui.adsabs.harvard.edu/abs/2010MNRAS.402.2049H} {402, 2049}

\bibitem[\protect\citeauthoryear{{Hey} et~al.,}{{Hey} et~al.}{2023}]{Hey2023}
{Hey} D.~R.,  et~al., 2023, \mn@doi [\aj] {10.3847/1538-3881/ad01bf}, \href {https://ui.adsabs.harvard.edu/abs/2023AJ....166..249H} {166, 249}

\bibitem[\protect\citeauthoryear{{Humphreys}}{{Humphreys}}{1978}]{Humphreys1978}
{Humphreys} R.~M.,  1978, \mn@doi [\apjs] {10.1086/190559}, \href {https://ui.adsabs.harvard.edu/abs/1978ApJS...38..309H} {38, 309}

\bibitem[\protect\citeauthoryear{{Jurcevic}, {Pierce}  \& {Jacoby}}{{Jurcevic} et~al.}{2000}]{Jurcevic2000}
{Jurcevic} J.~S.,  {Pierce} M.~J.,   {Jacoby} G.~H.,  2000, \mn@doi [\mnras] {10.1046/j.1365-8711.2000.03292.x}, \href {https://ui.adsabs.harvard.edu/abs/2000MNRAS.313..868J} {313, 868}

\bibitem[\protect\citeauthoryear{{Katz} et~al.,}{{Katz} et~al.}{2023}]{Katz2023}
{Katz} D.,  et~al., 2023, \mn@doi [\aap] {10.1051/0004-6361/202244220}, \href {https://ui.adsabs.harvard.edu/abs/2023A&A...674A...5K} {674, A5}

\bibitem[\protect\citeauthoryear{{Kiss}, {Szab{\'o}}  \& {Bedding}}{{Kiss} et~al.}{2006}]{Kiss2006}
{Kiss} L.~L.,  {Szab{\'o}} G.~M.,   {Bedding} T.~R.,  2006, \mn@doi [\mnras] {10.1111/j.1365-2966.2006.10973.x}, \href {https://ui.adsabs.harvard.edu/abs/2006MNRAS.372.1721K} {372, 1721}

\bibitem[\protect\citeauthoryear{{Lee} et~al.,}{{Lee} et~al.}{2011}]{Lee2011}
{Lee} J.~C.,  et~al., 2011, \mn@doi [\apjs] {10.1088/0067-0049/192/1/6}, \href {https://ui.adsabs.harvard.edu/abs/2011ApJS..192....6L} {192, 6}

\bibitem[\protect\citeauthoryear{{Levesque}}{{Levesque}}{2017}]{Levesque2017}
{Levesque} E.~M.,  2017, {Astrophysics of Red Supergiants}, \mn@doi{10.1088/978-0-7503-1329-2.
}

\bibitem[\protect\citeauthoryear{{Lomb}}{{Lomb}}{1976}]{Lomb1976}
{Lomb} N.~R.,  1976, \mn@doi [\apss] {10.1007/BF00648343}, \href {https://ui.adsabs.harvard.edu/abs/1976Ap&SS..39..447L} {39, 447}

\bibitem[\protect\citeauthoryear{{Massey}}{{Massey}}{2002}]{Massey2002}
{Massey} P.,  2002, \mn@doi [\apjs] {10.1086/338286}, \href {https://ui.adsabs.harvard.edu/abs/2002ApJS..141...81M} {141, 81}

\bibitem[\protect\citeauthoryear{{Massey}}{{Massey}}{2013}]{Massey2013}
{Massey} P.,  2013, \mn@doi [\nar] {10.1016/j.newar.2013.05.002}, \href {https://ui.adsabs.harvard.edu/abs/2013NewAR..57...14M} {57, 14}

\bibitem[\protect\citeauthoryear{{Massey} \& {Evans}}{{Massey} \& {Evans}}{2016}]{Massey2016}
{Massey} P.,  {Evans} K.~A.,  2016, \mn@doi [\apj] {10.3847/0004-637X/826/2/224}, \href {https://ui.adsabs.harvard.edu/abs/2016ApJ...826..224M} {826, 224}

\bibitem[\protect\citeauthoryear{{Massey}, {Levesque}, {Plez}  \& {Olsen}}{{Massey} et~al.}{2008}]{Massey2008}
{Massey} P.,  {Levesque} E.~M.,  {Plez} B.,   {Olsen} K. A.~G.,  2008, in {Bresolin} F.,  {Crowther} P.~A.,   {Puls} J.,  eds,  IAU Symposium Vol. 250, Massive Stars as Cosmic Engines. pp 97--110 (\mn@eprint {arXiv} {0801.1806}), \mn@doi{10.1017/S1743921308020383}

\bibitem[\protect\citeauthoryear{{Mathur} et~al.,}{{Mathur} et~al.}{2011}]{Mathur2011}
{Mathur} S.,  et~al., 2011, \mn@doi [\apj] {10.1088/0004-637X/741/2/119}, \href {https://ui.adsabs.harvard.edu/abs/2011ApJ...741..119M} {741, 119}

\bibitem[\protect\citeauthoryear{{Messineo}}{{Messineo}}{2023}]{Messineo2023}
{Messineo} M.,  2023, \mn@doi [\aap] {10.1051/0004-6361/202245587}, \href {https://ui.adsabs.harvard.edu/abs/2023A&A...671A.148M} {671, A148}

\bibitem[\protect\citeauthoryear{{Messineo} \& {Brown}}{{Messineo} \& {Brown}}{2019}]{Messineo2019}
{Messineo} M.,  {Brown} A.~G.~A.,  2019, \mn@doi [\aj] {10.3847/1538-3881/ab1cbd}, \href {https://ui.adsabs.harvard.edu/abs/2019AJ....158...20M} {158, 20}

\bibitem[\protect\citeauthoryear{{Messineo}, {Zhu}, {Menten}, {Ivanov}, {Figer}, {Kudritzki}  \& {Chen}}{{Messineo} et~al.}{2017}]{Messineo2017}
{Messineo} M.,  {Zhu} Q.,  {Menten} K.~M.,  {Ivanov} V.~D.,  {Figer} D.~F.,  {Kudritzki} R.-P.,   {Chen} C. H.~R.,  2017, \mn@doi [\apj] {10.3847/1538-4357/836/1/65}, \href {https://ui.adsabs.harvard.edu/abs/2017ApJ...836...65M} {836, 65}

\bibitem[\protect\citeauthoryear{{Mosenkov}, {Savchenko}, {Smirnov}  \& {Camps}}{{Mosenkov} et~al.}{2021}]{Mosenkov2021}
{Mosenkov} A.~V.,  {Savchenko} S.~S.,  {Smirnov} A.~A.,   {Camps} P.,  2021, \mn@doi [\mnras] {10.1093/mnras/stab2445}, \href {https://ui.adsabs.harvard.edu/abs/2021MNRAS.507.5246M} {507, 5246}

\bibitem[\protect\citeauthoryear{{Mosser} \& {Appourchaux}}{{Mosser} \& {Appourchaux}}{2009}]{Mosser2009}
{Mosser} B.,  {Appourchaux} T.,  2009, \mn@doi [\aap] {10.1051/0004-6361/200912944}, \href {https://ui.adsabs.harvard.edu/abs/2009A&A...508..877M} {508, 877}

\bibitem[\protect\citeauthoryear{{Mosser} et~al.,}{{Mosser} et~al.}{2011}]{Mosser2011}
{Mosser} B.,  et~al., 2011, \mn@doi [\aap] {10.1051/0004-6361/201015440}, \href {https://ui.adsabs.harvard.edu/abs/2011A&A...525L...9M} {525, L9}

\bibitem[\protect\citeauthoryear{{Ren} \& {Jiang}}{{Ren} \& {Jiang}}{2020}]{Ren2020}
{Ren} Y.,  {Jiang} B.-W.,  2020, \mn@doi [\apj] {10.3847/1538-4357/ab9c17}, \href {https://ui.adsabs.harvard.edu/abs/2020ApJ...898...24R} {898, 24}

\bibitem[\protect\citeauthoryear{{Ren}, {Jiang}, {Yang}  \& {Gao}}{{Ren} et~al.}{2019}]{Ren2019}
{Ren} Y.,  {Jiang} B.-W.,  {Yang} M.,   {Gao} J.,  2019, \mn@doi [\apjs] {10.3847/1538-4365/ab0825}, \href {https://ui.adsabs.harvard.edu/abs/2019ApJS..241...35R} {241, 35}

\bibitem[\protect\citeauthoryear{{Ren}, {Jiang}, {Yang}, {Wang}, {Jian}  \& {Ren}}{{Ren} et~al.}{2021a}]{Ren2021a}
{Ren} Y.,  {Jiang} B.,  {Yang} M.,  {Wang} T.,  {Jian} M.,   {Ren} T.,  2021a, \mn@doi [\apj] {10.3847/1538-4357/abcda5}, \href {https://ui.adsabs.harvard.edu/abs/2021ApJ...907...18R} {907, 18}

\bibitem[\protect\citeauthoryear{{Ren}, {Jiang}, {Yang}, {Wang}  \& {Ren}}{{Ren} et~al.}{2021b}]{Ren2021b}
{Ren} Y.,  {Jiang} B.,  {Yang} M.,  {Wang} T.,   {Ren} T.,  2021b, \mn@doi [\apj] {10.3847/1538-4357/ac307b}, \href {https://ui.adsabs.harvard.edu/abs/2021ApJ...923..232R} {923, 232}

\bibitem[\protect\citeauthoryear{{Scargle}}{{Scargle}}{1982}]{Scargle1982}
{Scargle} J.~D.,  1982, \mn@doi [\apj] {10.1086/160554}, \href {https://ui.adsabs.harvard.edu/abs/1982ApJ...263..835S} {263, 835}

\bibitem[\protect\citeauthoryear{{Sesar} et~al.,}{{Sesar} et~al.}{2010}]{Sesar2010}
{Sesar} B.,  et~al., 2010, \mn@doi [\apj] {10.1088/0004-637X/708/1/717}, \href {https://ui.adsabs.harvard.edu/abs/2010ApJ...708..717S} {708, 717}

\bibitem[\protect\citeauthoryear{{Skrutskie} et~al.,}{{Skrutskie} et~al.}{2006}]{Skrutskie2006}
{Skrutskie} M.~F.,  et~al., 2006, \mn@doi [\aj] {10.1086/498708}, \href {https://ui.adsabs.harvard.edu/abs/2006AJ....131.1163S} {131, 1163}

\bibitem[\protect\citeauthoryear{{Soraisam} et~al.,}{{Soraisam} et~al.}{2018}]{Soraisam2018}
{Soraisam} M.~D.,  et~al., 2018, \mn@doi [\apj] {10.3847/1538-4357/aabc59}, \href {https://ui.adsabs.harvard.edu/abs/2018ApJ...859...73S} {859, 73}

\bibitem[\protect\citeauthoryear{{Soszy{\'n}ski} et~al.,}{{Soszy{\'n}ski} et~al.}{2013}]{Soszy2013}
{Soszy{\'n}ski} I.,  et~al., 2013, \mn@doi [\actaa] {10.48550/arXiv.1304.2787}, \href {https://ui.adsabs.harvard.edu/abs/2013AcA....63...21S} {63, 21}

\bibitem[\protect\citeauthoryear{{Stencel}, {Ueta}, {Wall}  \& {Yamamura}}{{Stencel} et~al.}{2009}]{Stencel2009}
{Stencel} R.~E.,  {Ueta} T.,  {Wall} R.~J.,   {Yamamura} I.,  2009, in {Onaka} T.,  {White} G.~J.,  {Nakagawa} T.,   {Yamamura} I.,  eds,  Astronomical Society of the Pacific Conference Series Vol. 418, AKARI, a Light to Illuminate the Misty Universe. p.~459

\bibitem[\protect\citeauthoryear{{Udalski}}{{Udalski}}{2003}]{Udalski2003}
{Udalski} A.,  2003, \mn@doi [\actaa] {10.48550/arXiv.astro-ph/0401123}, \href {https://ui.adsabs.harvard.edu/abs/2003AcA....53..291U} {53, 291}

\bibitem[\protect\citeauthoryear{{Udalski}, {Kubiak}  \& {Szymanski}}{{Udalski} et~al.}{1997}]{Udalski1997}
{Udalski} A.,  {Kubiak} M.,   {Szymanski} M.,  1997, \mn@doi [\actaa] {10.48550/arXiv.astro-ph/9710091}, \href {https://ui.adsabs.harvard.edu/abs/1997AcA....47..319U} {47, 319}

\bibitem[\protect\citeauthoryear{{Udalski}, {Szyma{\'n}ski}  \& {Szyma{\'n}ski}}{{Udalski} et~al.}{2015}]{Udalski2015}
{Udalski} A.,  {Szyma{\'n}ski} M.~K.,   {Szyma{\'n}ski} G.,  2015, \mn@doi [\actaa] {10.48550/arXiv.1504.05966}, \href {https://ui.adsabs.harvard.edu/abs/2015AcA....65....1U} {65, 1}

\bibitem[\protect\citeauthoryear{{Van Dyk}}{{Van Dyk}}{2017}]{VanDyk2017}
{Van Dyk} S.~D.,  2017, \mn@doi [Philosophical Transactions of the Royal Society of London Series A] {10.1098/rsta.2016.0277}, \href {https://ui.adsabs.harvard.edu/abs/2017RSPTA.37560277V} {375, 20160277}

\bibitem[\protect\citeauthoryear{{Wen}, {Gao}, {Yang}, {Chen}, {Ren}, {Wang}  \& {Jiang}}{{Wen} et~al.}{2024}]{Wen2024}
{Wen} J.,  {Gao} J.,  {Yang} M.,  {Chen} B.,  {Ren} Y.,  {Wang} T.,   {Jiang} B.,  2024, \mn@doi [\aj] {10.3847/1538-3881/ad12bf}, \href {https://ui.adsabs.harvard.edu/abs/2024AJ....167...51W} {167, 51}

\bibitem[\protect\citeauthoryear{{Yang} \& {Jiang}}{{Yang} \& {Jiang}}{2011}]{Yang2011}
{Yang} M.,  {Jiang} B.~W.,  2011, \mn@doi [\apj] {10.1088/0004-637X/727/1/53}, \href {https://ui.adsabs.harvard.edu/abs/2011ApJ...727...53Y} {727, 53}

\bibitem[\protect\citeauthoryear{{Yang} \& {Jiang}}{{Yang} \& {Jiang}}{2012}]{Yang2012}
{Yang} M.,  {Jiang} B.~W.,  2012, \mn@doi [\apj] {10.1088/0004-637X/754/1/35}, \href {https://ui.adsabs.harvard.edu/abs/2012ApJ...754...35Y} {754, 35}

\bibitem[\protect\citeauthoryear{{Yang} et~al.,}{{Yang} et~al.}{2019}]{Yang2019}
{Yang} M.,  et~al., 2019, \mn@doi [\aap] {10.1051/0004-6361/201935916}, \href {https://ui.adsabs.harvard.edu/abs/2019A&A...629A..91Y} {629, A91}

\bibitem[\protect\citeauthoryear{{Yuan}, {Liu}  \& {Xiang}}{{Yuan} et~al.}{2013}]{Yuan2013}
{Yuan} H.~B.,  {Liu} X.~W.,   {Xiang} M.~S.,  2013, \mn@doi [\mnras] {10.1093/mnras/stt039}, \href {https://ui.adsabs.harvard.edu/abs/2013MNRAS.430.2188Y} {430, 2188}

\bibitem[\protect\citeauthoryear{{Zhang}, {Green}  \& {Rix}}{{Zhang} et~al.}{2023}]{Zhang2023}
{Zhang} X.,  {Green} G.~M.,   {Rix} H.-W.,  2023, \mn@doi [\mnras] {10.1093/mnras/stad1941}, \href {https://ui.adsabs.harvard.edu/abs/2023MNRAS.524.1855Z} {524, 1855}

\bibitem[\protect\citeauthoryear{{Zhang}, {Ren}, {Jiang}, {Soszynski}  \& {Jayasinghe}}{{Zhang} et~al.}{2024}]{Zhang2024}
{Zhang} Z.,  {Ren} Y.,  {Jiang} B.,  {Soszynski} I.,   {Jayasinghe} T.,  2024, \mn@doi [arXiv e-prints] {10.48550/arXiv.2405.01405}, \href {https://ui.adsabs.harvard.edu/abs/2024arXiv240501405Z} {p. arXiv:2405.01405}

\makeatother
\end{thebibliography}




\appendix
\section{Completeness and Purity} \label{sec:completeness and purity}
\renewcommand{\thefigure}{A\arabic{figure}}
\setcounter{figure}{0}
To roughly evaluate the performance of the classification method, we calculate two parameters, recall and precision, both obtained from the confusion matrix. We use these two parameters to characterize the completeness and purity of the RSG candidates obtained in the  removing AGBs step of our method. The confusion matrix for the reference sample stars is shown in Fig.~\ref{fig:Consusion matrix}. As observed, according to the granulation parameters, 962 RSGs are correctly classified, and 599 non-RSGs are incorrectly classified as RSGs. The precision and recall are approximately 62\% and 84\%, respectively. We note that in our following analysis, we have further eliminated the contamination of the RGBs from these RSG candidates obtained in the previous step, thus improving the precision of our final sample. However, due to the incompleteness of Gaia data and lack of the training samples, we can not accurately estimate the completeness and precision of our sample.

\begin{figure}
    \centering
    \includegraphics[width=0.45\textwidth]{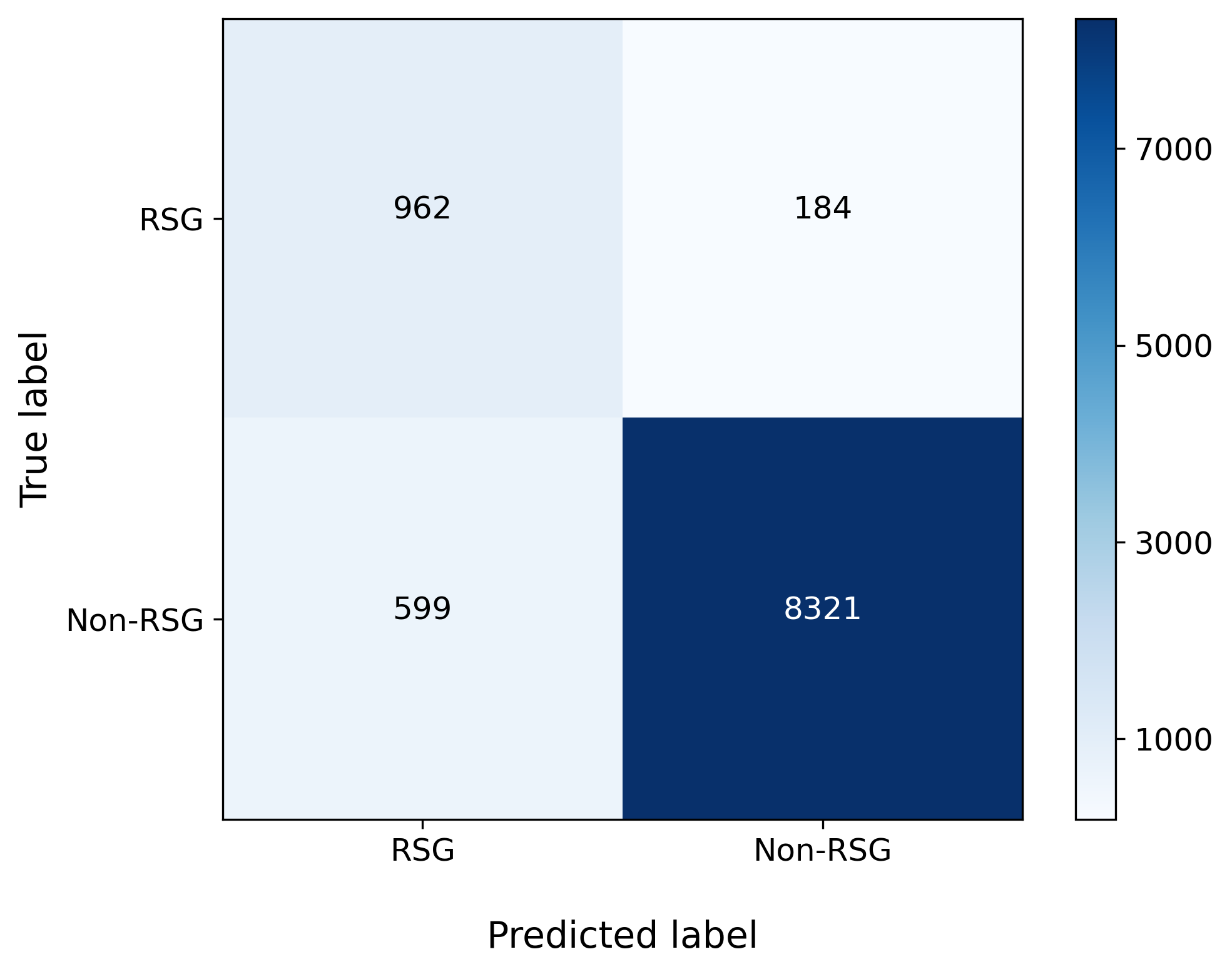}
    \caption{The confusion matrix of visual classification results for the reference sample stars in the $\tau_{\rm{eff}}$ $-$ $\sigma_{\rm{gran}}$ diagram.}
    \label{fig:Consusion matrix}
\end{figure}

In the previous research, \citet{Messineo2019} showed a sample of 889 nearby RSG candidates, which is the largest RSG sample in the Milky Way so far. In practice, even if we consider the stars in their Area A and B, which generally have $M_{\rm bol} \textless -5 $  mag and are considered RSGs, the precision is only about 41\%. In the current work, the precision can reach more than 62\%. In sum, precision has a great improvement. 

\section{Distribution of Sample Luminosities: Sample Limitations} \label{sec:distribution of sample luminosities: sample limitations}
\renewcommand{\thefigure}{B\arabic{figure}}
\setcounter{figure}{0}
As shown in Fig.~\ref{fig:training_sample}, the characteristic amplitudes of RSG candidates cover a very small range. According to the study by \citet{Ren2020}, the amplitude of granulation increases with luminosity. Thus we analyse the limitation of characteristic amplitudes from the perspective of luminosity. We examine the magnitude distribution of the RSGs from reference sample and find that the magnitudes are concentrated at the faint end, ranging from $-$8 to $-$6\,mag. Therefore, we speculate the method may only be suitable for identifying faint RSGs. To verify this hypothesis, we search for bright RSGs in the M31 from \citet{Ren2020}. The magnitudes of sample stars from \citet{Ren2020} almost all distribute between $-$11 and $-$8\,mag, which are obviously brighter than reference sample stars, as shown in Fig.~\ref{fig:compare_magnitude}. Besides, the median of characteristic amplitudes of their RSGs is 64.03\,mmag, which is significantly larger than our selection criteria. We also present the magnitude distribution of the new RSG sample stars, as shown in red stripes in Fig.~\ref{fig:compare_magnitude}, which clearly shows that the magnitudes are concentrated at the faint end. Based on the above analysis, we conclude that the method is indeed only applicable to distinguish the faint RSGs from AGBs.

\begin{figure}
\centering
\includegraphics[width=0.45\textwidth]{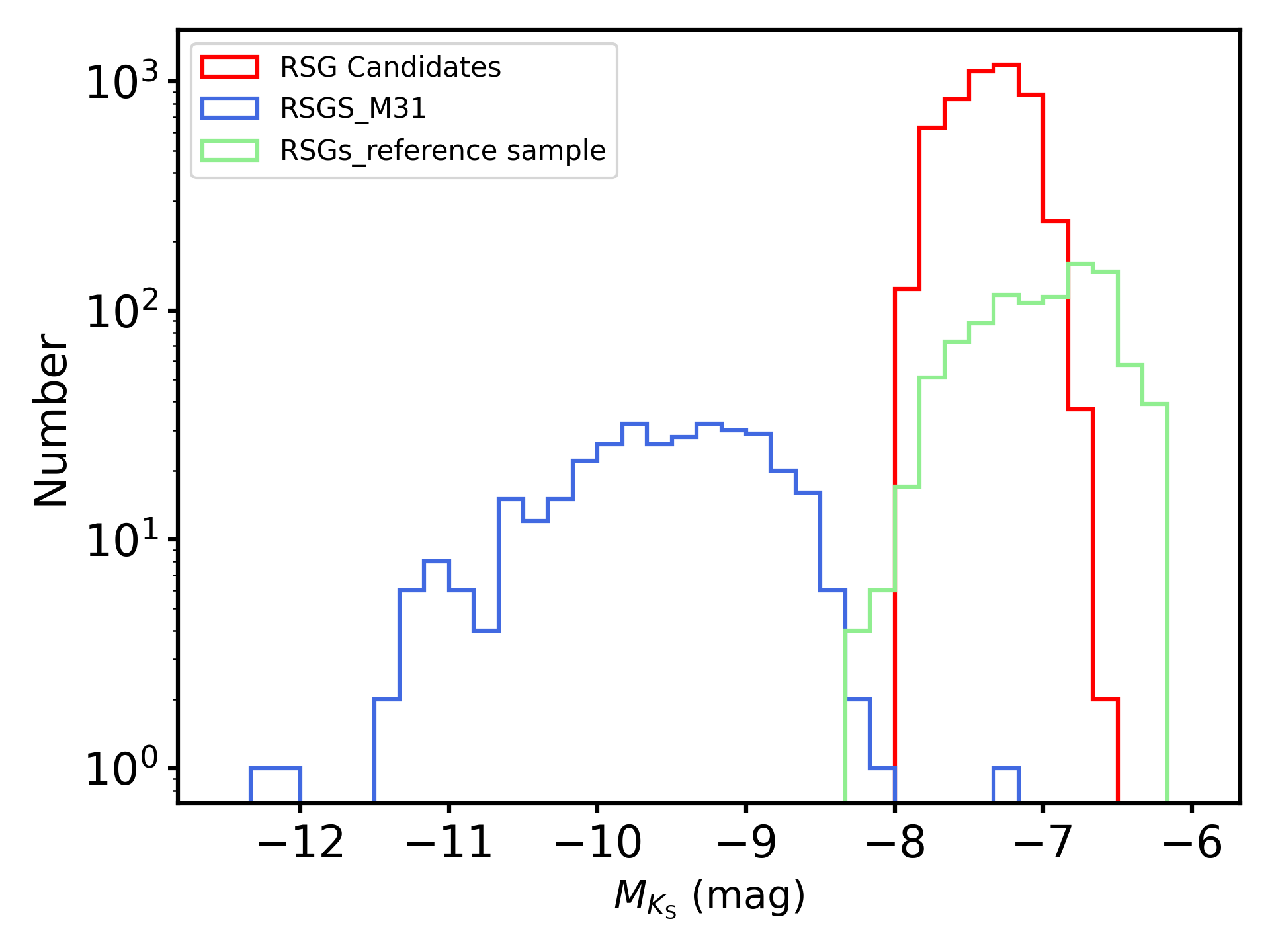}
\caption{The magnitude comparisons between RSG candidates, the RSGs from our reference sample, and the RSGs from \citet{Ren2020} are presented. Red stripes indicate the RSG candidates in this work and green stripes represent RSGs in the reference sample, respectively. Blue stripes show the RSGs in M31 from \citet{Ren2020}.}
\label{fig:compare_magnitude}
\end{figure}

Moreover, in addition to the different luminosities leading to differences in granulation amplitudes, the observational bands may also contribute to discrepancies in characteristic amplitudes. The time-series data from the iPTF survey $R$-band are utilised to determine the granulation parameters in the study by \citet{Ren2020}. The light curves from the OGLE survey $I$-band are utilised to obtain the granulation parameters in this work. However, the mechanism behind the discrepancy in granulation characteristic amplitudes caused by different observational bands remains unclear.

To sum up, although the visual classification method has relatively high precision and recall, this method still has a limitation. The method is only effective for identifying faint RSGs, whose characteristic amplitudes are obviously smaller than those of AGBs.

\section{Determining of distances of RSG candidates} \label{sec:Determining of distances of RSG candidates}
\renewcommand{\thefigure}{C\arabic{figure}}
\setcounter{figure}{0}
To investigate the distribution of new RSG sample in the Milky Way, we then determine the distances to the RSG candidates. Initially we fit the relation between $K_{\rm S}$-band absolute magnitude and $\sigma_{\rm{gran}}$, as well as the relation between intrinsic color and $\sigma_{\rm{gran}}$ of the RSGs in the reference sample, as shown in Fig.~\ref{fig:magnitude-color-sigma}. The fitting errors, respectively, are 0.272 and 0.067\,mag. For each relation, we perform two fits, applying a 3$\sigma$ clipping to remove outliers that may affect the fitting results during the second fit. The extinction is corrected using the reddening maps of MCs and the reddening coefficients from \citet{Chen2022}. Additionally, we assume the distances to LMC and SMC are 50\,kpc and 60\,kpc \citep{Lee2011}, respectively. Finally, the fitting relations are expressed in the following form:

\begin{equation}
\begin{aligned}
     (J-K_{\rm S})_{0} &= 0.036 \sigma_{\rm{gran}} + 0.791,  \\
     M_{K_{\rm S}} &= -0.301 \sigma_{\rm{gran}} - 6.458.  
\label{equ:the fitting relation}
\end{aligned}
\end{equation}

Then, we apply these relations to the final RSG candidates and obtain the $K_{\rm S}$-band absolute magnitude and intrinsic color. Consequently, the distances of the new Galactic RSGs are derived by $d = 10^{\frac{K_{\rm S} - M_{K_{\rm S}} - A_{K_{\rm S}} + 5}{5}}$. The $K_{\rm S}$-band extinction is calculated by $\frac{R_{K_{\rm S}}}{R_{J} - R_{K_{\rm S}}} \times E(J - K_{\rm S})$, with the extinction coefficients adopted from \citet{Yuan2013}.

\begin{figure}
\includegraphics[width=0.45\textwidth]{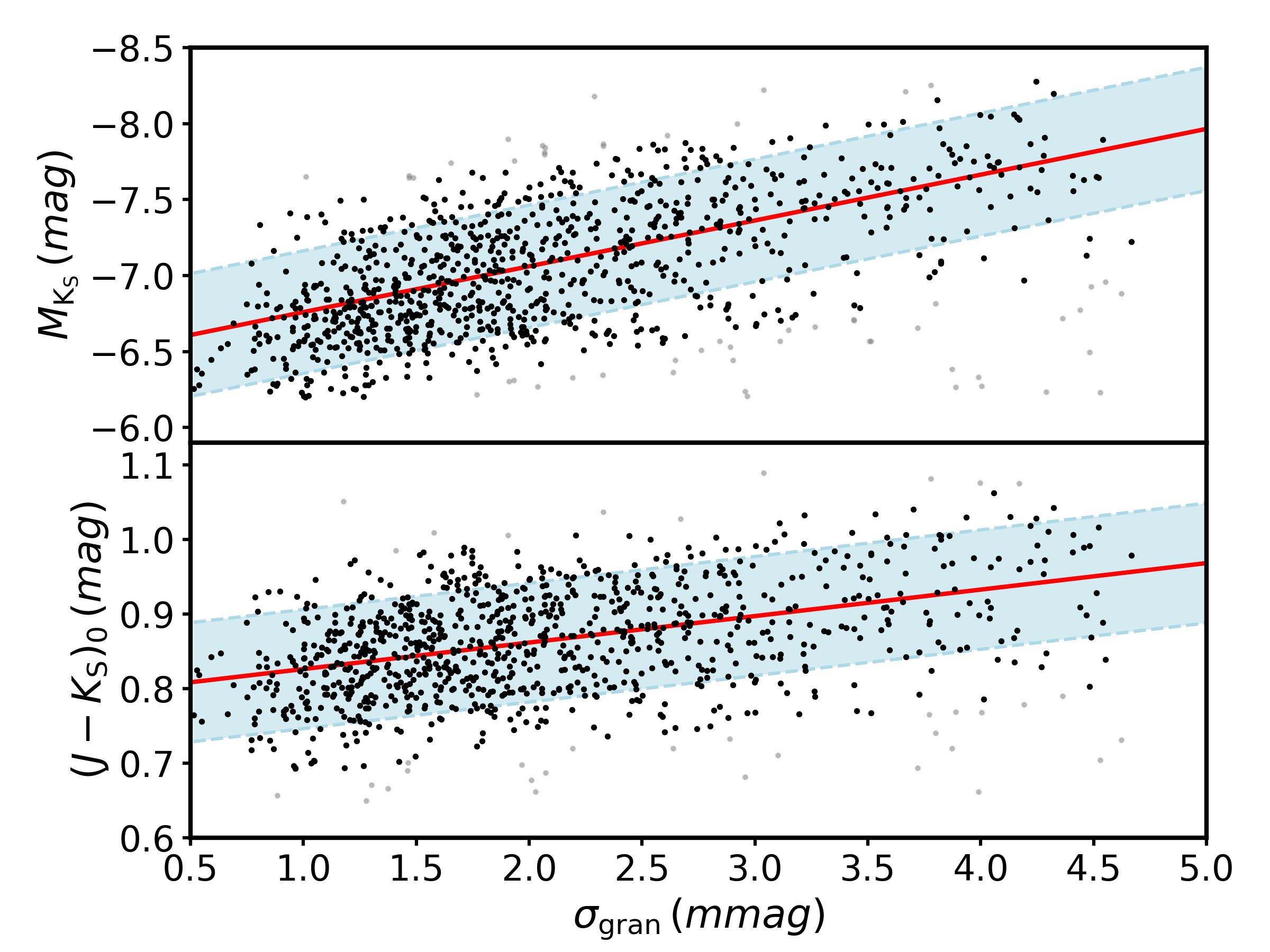}
\caption{The upper panel is the relation between $K_{\rm S}$-band absolute magnitude and $\sigma_{\rm{gran}}$ for the RSGs in the reference sample. The grey open circles represent outliers that have been discarded. The red line indicates the linear fit. The blue dashed lines show the linear fit with the 95\% confidence region (blue shaded region). The lower panel is the same as the upper panel, but for the relation between intrinsic color $(J-K_{\rm S})_{0}$ and $\sigma_{\rm{gran}}$.}
\label{fig:magnitude-color-sigma}
\end{figure}

\section{Distance Uncertainty} \label{sec:distance uncertainty}
Such a large sample of RSGs can help us study the structure of our Galaxy. Thus, we estimate the distance uncertainty and we take error in the apparent magnitude, absolute magnitude and the extinction into account. In this study, we use the median error value for the $J$ and $K_{\rm S}$ band, 0.029 and 0.026\,mag, respectively, as the typical apparent magnitude errors. The absolute magnitude is obtained by the fitting the relation of absolute magnitude and $\sigma_{\rm{gran}}$, with a dispersion of 0.272\,mag. The extinction error is calculated based on the apparent magnitude error in the $J$ and $K_{\rm S}$ bands and the fitting error of intrinsic color, given by $\frac{R_{K_{\rm S}}}{R_{J} - R_{K_{\rm S}}} \times \sqrt{0.026^{2} + 0.029^{2} + 0.067^{2}} \approx 0.057$\,mag, with the extinction coefficients adopted from \citet{Yuan2013}. Therefore, the dispersion of $\mu$ is calculated by $\sigma_{\mu} = \sqrt{0.026^{2} + 0.272^{2} + 0.057^{2}} \approx 0.279$\,mag. Then, we obtained the final distance uncertainty, given as $D \times \rm ln(10) \times 0.2\sigma_{\mu}$ and the typical error is approximately 13\%.

\bsp	
\label{lastpage}
\end{document}